\documentstyle[12pt]{article}
\begin{document}
\centerline{\Large\bf Born-Infeld-Einstein theory with matter}
\vskip .7in
\centerline{Dan N. Vollick}
\centerline{Department of Physics and Astronomy}
\centerline{University of British Columbia Okanagan}
\centerline{3333 University Way}
\centerline{Kelowna, B.C.}
\centerline{V1V 1V7}
\vskip .9in
\centerline{\bf\large Abstract}
\vskip 0.5in
The field equations associated with the Born-Infeld-Einstein action 
including matter are derived using a Palatini 
variational principle. Scalar, electromagnetic, and Dirac fields 
are considered. It is shown that an action can be chosen for the scalar field
that produces field equations identical to the usual Einstein field equations
minimally coupled to a scalar field. In the electromagnetic and Dirac cases
the field equations reproduce the standard equations only to lowest order.
The spherically symmetric electrovac equations are 
studied in detail. It is shown that the resulting Einstein equations 
correspond to gravity coupled to a modified Born-Infeld theory. It is also
shown that point charges are not allowed. All particles must have a finite
size. Mass terms for the fields are also considered.
\newpage
\section*{Introduction}
In the 1930's Born and Infeld \cite{Bo1} attempted to eliminate 
the divergent self energy of the electron by modifying Maxwell's
theory.
Born-Infeld electrodynamics follows from the Lagrangian 
\begin{equation}
L=-\frac{1}{4\pi b}\left\{\sqrt{-det(g_{\mu\nu}+bF_{\mu\nu})}
-\sqrt{-det(g_{\mu\nu})}\right\}\; ,
\end{equation}
where $g_{\mu\nu}$ is the metric tensor and $F_{\mu\nu}$ is the
electromagnetic field tensor.
In the weak field limit this Lagrangian reduces to the
Maxwell Lagrangian plus small corrections. For strong fields the
field equations deviate significantly from Maxwell's theory and 
the self energy of the electron can be shown to be finite.
The Born-Infeld action also appears in string theory. The action
for a D-brane is of the Born-Infeld form with two fields, a
gauge field on the brane and the projection of the Neveu-Schwarz
B-field onto the brane \cite{Po1}.
   
In a recent paper \cite{Vo1} I considered using a Palatini variational
approach to derive the field equations associated with the action
\begin{equation}
L=-\frac{1}{\kappa b}\left\{\sqrt{-det(g_{\mu\nu}+bR_{\mu\nu})}
-\sqrt{-det(g_{\mu\nu})}\right\},
\label{intro}
\end{equation}
where $R_{\mu\nu}$ is the Ricci tensor,
and $\kappa=8\pi G$. This action has also been examined using a purely metric
variation by Deser and Gibbons \cite{De1}, Feigenbaum, Freund and Pigli \cite{Fe1} and Feigenbaum \cite{Fe2}.

In this paper I will consider adding matter to
the theory by including it in the determinant in equation (\ref{intro}).
The forms of matter that I investigate in this paper include scalar, 
electromagnetic, and
Dirac fields. In all the cases except for the scalar field there are corrections
to the standard equations. In the scalar field case it is shown that it is
possible to chose a Lagrangian that gives the standard equations exactly.
The fields produced by a spherically symmetric charge distribution are also
studied. It is shown that the resulting Einstein equations 
correspond to gravity coupled to a modified Born-Infeld theory and
that point charges are not allowed in the theory.
 
\section*{The Field Equations} 
The field equations for the theory follow from the Born-Infeld-Einstein
action
\begin{equation}
L=-\frac{1}{\kappa b}\left\{\sqrt{-det(g_{\mu\nu}+
bR_{\mu\nu}+\kappa bM_{\mu\nu})}-\sqrt{-det(g_{\mu\nu})}\right\}\;,
\label{lag}
\end{equation}
where $R_{\mu\nu}$ is the Ricci tensor, $\kappa=8\pi G$, $b$ 
is a constant, 
and $M_{\mu\nu}$ is the matter contribution. For a simple example of 
the matter contribution
consider a scalar field $\phi$. In this case $M_{\mu\nu}= \nabla_{\mu}
\phi\nabla_{\nu}\phi$. The Ricci tensor is given by
\begin{equation}
R_{\mu\nu}=\partial_{\nu}\Gamma^{\alpha}_{\mu\alpha}-\partial_{\alpha}
\Gamma^{\alpha}_{\mu\nu}-\Gamma^{\alpha}_{\beta\alpha}\Gamma^{\beta}_{\mu\nu}
+\Gamma^{\alpha}_{\beta\mu}\Gamma^{\beta}_{\alpha\nu}
\label{Ricci1}
\end{equation}
and the connection is taken to be symmetric. Note that $R_{\mu\nu}$ is not
symmetric in general. 
   
If the curvature and matter terms are small an 
expansion of (\ref{lag}) gives
\begin{equation}
L\simeq -\sqrt{g}\left(\frac{1}{2\kappa}R+\frac{1}{2}M\right)\;.
\label{approx}
\end{equation}
Thus, if $M$ does not vanish, the lowest order contribution to the 
matter Lagrangian is
\begin{equation}
L_M\simeq -\frac{1}{2}\sqrt{g}M.
\label{approxLag}
\end{equation}
In the case of the electromagnetic field $M$ vanishes, so higher order terms
will be needed.
   
Varying the action with respect to $g_{\mu\nu}$ gives
\begin{equation}
\sqrt{P}\left(P^{-1}\right)^{(\mu\nu)}+\kappa b\sqrt{P}\left(P^{-1}\right)^
{\alpha\beta}\frac{\partial M_{\beta\alpha}}{\partial g_{\mu\nu}}=\sqrt{g}g^{\mu\nu}
\label{eq1}\; ,
\end{equation}
where $P_{\mu\nu}=g_{\mu\nu}+bR_{\mu\nu}+\kappa bM_{\mu\nu}$, $P^{-1}$ is the inverse of $P$, $(P^{-1})^{(\mu\nu)}$
is the symmetric part of $P^{-1}$, $P=-det(P_{\mu\nu})$ and $g=-det(g_{\mu\nu})$.
If $M_{\mu\nu}$ is independent of the metric, as will be the case for massless
scalar and electromagnetic fields, the second term on the left hand side
in equation (\ref{eq1}) will vanish and the field equations will simplify
significantly.
 
Varying with respect to $\Gamma^{\alpha}_{\mu\nu}$ gives
\begin{equation}
\nabla_{\alpha}\left[\sqrt{P}\left(P^{-1}\right)^{(\mu\nu)}\right]
-\frac{1}{2}\nabla_{\beta}\left\{\sqrt{P}\left[\delta^{\mu}_{\alpha}
\left(P^{-1}\right)^{\beta\nu}+\delta^{\nu}_{\alpha}\left(P^{-1}
\right)^{\beta\mu}\right]\right\}=0\; .
\label{eq2}
\end{equation}
Here I have assumed that $M_{\mu\nu}$ is independent of the connection. This
will be true for scalar and electromagnetic fields, but not for the Dirac
field. The Dirac field will be studied in detail later.
Contracting over $\alpha$ and $\nu$ gives
\begin{equation}
\nabla_{\alpha}\left[\sqrt{P}\left(P^{-1}\right)^{[\alpha\mu]}
\right]=-\frac{3}{5}\nabla_{\alpha}\left[\sqrt{P}\left(P^{-1}
\right)^{(\alpha\mu)}\right]\;,
\label{anti}
\end{equation}
where $(P^{-1})^{[\mu\nu]}$ is the antisymmetric part of $P^{-1}$.
  
From equation (\ref{anti}) we see that
\begin{equation}
\nabla_{\alpha}\left[\sqrt{P}\left(P^{-1}\right)^{\alpha\mu}\right]=
\frac{2}{5}\nabla_{\alpha}\left[\sqrt{P}\left(P^{-1}\right)^
{(\alpha\mu)}\right]\; .
\label{eq2b}
\end{equation}
Taking $M_{\mu\nu}$ to be independent of the metric and
substituting equations (\ref{eq1}) and (\ref{eq2b}) into (\ref{eq2}) gives
\begin{equation}
\nabla_{\alpha}\left[\sqrt{g}g^{\mu\nu}\right]-\frac{1}{5}\nabla_{\beta}\left\{
\sqrt{g}\left[g^{\mu\beta}\delta^{\nu}_{\alpha}+g^{\nu\beta}\delta^{\mu}_
{\alpha}\right]\right\}=0.
\label{last}
\end{equation}
Since the trace of the above system of equations vanishes there are four
too few equations and the system is under determined. Thus, we expect four
arbitrary functions in the solution. Such a solution is given by
\begin{equation}
\nabla_{\alpha}\left[\sqrt{g}g^{\mu\nu}\right]=
\sqrt{g}\left[\delta^{\mu}_{\alpha}V^{\nu}+\delta^{\nu}_{\alpha}
V^{\mu}\right]\;,
\label{matter}
\end{equation}
where $V^{\mu}$ is an arbitrary vector. The connection that follows from
this set of equations is given by \cite{Vo1}
\begin{equation}
\Gamma^{\alpha}_{\mu\nu}=\left\{
\begin{array}{ll}
\alpha\\
\mu\nu\\
\end{array}
\right\}+\frac{1}{2}\left[3g_{\mu\nu}V^{\alpha}-\delta^{\alpha}_{\mu}V_{\nu}-
\delta^{\alpha}_{\nu}V_{\mu}\right]\; ,
\label{connection}
\end{equation}
where the first term on the right hand side is the Christoffel symbol.
This new ``geometrical" vector field was discussed in detail in \cite{Vo1} and will not
be considered here. I will therefore set $V_{\mu}=0$ and the connection 
is the Christoffel symbol. 
  
Thus, the gravitational field equations are given by (\ref{eq1}) with the connection given by the Christoffel symbol if $M_{\mu\nu}$ is independent
of the metric. The matter field equations are
given by varying the action with respect to the matter fields. I will study
these equations in the subsequent sections, once a specific type of matter
field has been chosen.

\section*{The Scalar Field}
For the scalar field I will take $M_{\mu\nu}=\nabla_{\mu}
\phi\nabla_{\nu}\phi$. In this case 
$M_{\mu\nu}$ is symmetric and is independent
of the metric. Note that $R_{\mu\nu}$ is symmetric since the connection is
just the Christoffel symbol. This implies that $P_{\mu\nu}$ is 
symmetric.
  
Consider the general case in which $M_{\mu\nu}$ is symmetric and independent
of the metric. Taking the 
determinant of equation (\ref{eq1}) gives $P=g$ and after some simple algebra
we find that
\begin{equation}
G_{\mu\nu}=-\kappa\left(M_{\mu\nu}-\frac{1}{2}Mg_{\mu\nu}\right)\;. 
\label{EFEa}
\end{equation}
Thus, for the scalar field we have
\begin{equation}
G_{\mu\nu}=-\kappa\left(\nabla_{\mu}\phi\nabla_{\nu}\phi-\frac{1}{2}
g_{\mu\nu}\nabla^{\alpha}\phi\nabla_{\alpha}\phi\;\right) .
\end{equation}
Note that these are the standard Einstein field equations minimally coupled
to a scalar field. 
 
The scalar field equation can be found by varying the action with respect
to $\phi$. The resulting field equation is
\begin{equation}
\nabla_{\mu}\left[\sqrt{P}\left(P^{-1}\right)^{\mu\nu}\nabla_{\nu}\phi\right]=0\;.
\end{equation}
Using (\ref{eq1}) gives the usual field equations
\begin{equation}
\nabla^{\alpha}\nabla_{\alpha}\phi=0 \; .
\end{equation}
Note that the lowest order matter Lagrangian 
given in equation (\ref{approxLag}) is the standard Lagrangian for a scalar field.
Therefore, the higher order terms in both the gravitational and matter 
Lagrangians do not modify the field equations.
This extends the universality of the Einstein vacuum equations
discovered by Ferraris et al. \cite{Fer1,Fer2} to Born-Infeld Einstein actions
with a scalar field. In their papers Ferraris et al. showed that 
Lagrangians of the form $L=F(R)$ and $L=f(R^{\mu\nu}R_{\mu\nu})$ always 
give the Einstein vacuum equations with a cosmological constant under a 
Palatini variation.
  
Other choices for $M_{\mu\nu}$ are possible. The most general expression
that is bilinear in $\nabla\phi$ is $M_{\mu\nu}=\nabla_{\mu}\phi\nabla_{\nu}\phi
+\alpha g_{\mu\nu}\nabla^{\alpha}\phi\nabla_{\alpha}\phi$, where $\alpha$
is a constant. Such choices lead
to more complicated theories in which the connection is no longer the Christoffel
symbol.

\section*{The Electromagnetic Field}
For the electromagnetic field take $M_{\mu\nu}=\alpha F_{\mu\nu}$, where
$\alpha$ is a constant. In this case
$M_{\mu\nu}$ is independent of the metric, but $P_{\mu\nu}$ is no longer symmetric.
This implies that the connection is the Christoffel symbol.
The field equations
are difficult to solve exactly, but if the fields are sufficiently weak 
they can be expanded
to second order in the fields. The following expressions, computed to quadratic
order in the field variables, will be useful in finding the field equations
\begin{equation}
\left(P^{-1}\right)^{(\mu\nu)}=g^{\mu\nu}-bR^{\mu\nu}+\alpha^2\kappa^2b^2F^{\mu}_{\;\;\;\alpha}F^{\alpha\nu}
+b^2R^{\mu}_{\;\;\alpha}R^{\alpha\nu}
\end{equation}
and
\begin{equation}
\sqrt{P}=\sqrt{g}\left[1+\frac{1}{2}bR+\frac{1}{8}b^2R^2+\frac{1}{4}\alpha^2\kappa^2
b^2F_{\alpha\beta}F^{\alpha\beta}
-\frac{1}{4}b^2R_{\alpha\beta}R^{\beta\alpha}\right]\; .
\end{equation}
The field equations up to and including quadratic terms in the fields are given by
\begin{equation}
G_{\mu\nu}=\frac{1}{8}b\left[R^2g_{\mu\nu}-4RR_{\mu\nu}-2g_{\mu\nu}R_{\alpha\beta}R^{\alpha\beta}
+8R^{\mu}_{\;\;\;\alpha}R^{\alpha\nu}\right]
-\alpha^2\kappa^2b\left[F_{\mu}^{\;\;\;\alpha}F_{\nu\alpha}-\frac{1}{4}g_{\mu\nu}
F^{\alpha\beta}F_{\alpha\beta}\right]\; .
\label{eqem}
\end{equation}
Note that electromagnetic terms on the right hand side correspond
to the energy-momentum tensor of the electromagnetic field.
Thus, to obtain the Einstein field equations to lowest order we must take
$\alpha^2=1/(\kappa b)$. This implies that $b$ must be positive.
Also note that there are correction terms to the
Einstein field equations that only vanish in the limit $b\rightarrow 0$.
In fact, certain cubic and quartic terms need to be included for the approximation
to be consistent. To see this note that $R\sim\kappa F^2$ which implies that
the quadratic corrections in (\ref{eqem})
are $\sim b(\kappa F^2)^2$. However, terms like $b\kappa RF^2$
and $b\kappa^2F^4$ are of the same order and therefore need to be included
in the approximation. Since the exact form of the corrections
is not important I will not consider these additional terms further.
  
The field equations for the electromagnetic field are derived by varying
the action with respect to $A_{\mu}$ and are given by
\begin{equation}
\nabla_{\mu}\left[\sqrt{P}\left(P^{-1}\right)^{[\mu\nu]}\right]=0\;.
\label{eq22a}
\end{equation}
To lowest order the above reduces to the usual field equations
\begin{equation}
\nabla_{\mu}F^{\mu\nu}=0\;.
\label{antisymm}
\end{equation}
  
The lowest order contribution to the electromagnetic field Lagrangian cannot
be computed from (\ref{approxLag}) since $M=0$. The Lagrangian to first order
in the curvature and second order in the electromagnetic field is given by
\begin{equation}
L\simeq-\sqrt{g}\left[\frac{1}{2\kappa}R+\frac{1}{4}F_{\alpha\beta}F^{\alpha\beta}
\right]\;.
\label{23}
\end{equation}
Thus, to this order, we obtain the standard Einstein-Maxwell Lagrangian.
  
It is interesting to note that the inclusion of the electromagnetic field
$F_{\mu\nu}$ excludes the possibility of the ``geometric" vector field
discussed earlier. To see
this note that (\ref{anti}) and (\ref{eq22a}) imply that
\begin{equation}
\nabla_{\alpha}\left[\sqrt{P}\left(P^{-1}\right)^{(\alpha\mu)}\right]=0\;.
\end{equation}
This together with (\ref{eq1}) and (\ref{last}) implies that
\begin{equation}
\nabla_{\alpha}\left[\sqrt{g}g^{\mu\nu}\right]=0\;,
\end{equation}
which tells us that the connection is the Christoffel symbol and that there
is no room for an extra ``geometric" vector field.

In general it seems that it is difficult to write the field equations
in the form $G_{\mu\nu}=-\kappa T_{\mu\nu}$. However, it is possible to write
them in this form for spherically symmetric fields. In this case we can take
\begin{equation}
A_{\mu}=\left(\phi,\vec{0}\right)\; ,
\end{equation}
where $\phi$ depends only on $r$ and the metric is diagonal. A short calculation
shows that
\begin{equation}
P_{\mu\nu}=\left[
\begin{array}{ll}
\;\;q_{00}\;\;\;\;\;\;\Phi^{'}\;\;\;\;\;0\;\;\;\;\;\;0\\
-\Phi^{'}\;\;\;\;\;q_{11}\;\;\;\;\;0\;\;\;\;\;\;0\\
\;\;\;0\;\;\;\;\;\;\;\;0\;\;\;\;\;q_{22}\;\;\;\;\;0\\
\;\;\;0\;\;\;\;\;\;\;\;0\;\;\;\;\;\;0\;\;\;\;\;q_{33}\\
\end{array}
\right]
\label{P}
\end{equation}
and
\begin{equation}
\left(P^{-1}\right)^{\mu\nu}=\left[
\begin{array}{ll}
\;\frac{q_{11}}{\Delta}\;\;-\frac{\Phi^{'}}{\Delta}\;\;\;\;\;\;0\;\;\;\;\;\;0\\
\\
\;\frac{\Phi^{'}}{\Delta}\;\;\;\;\;\;\;\frac{q_{00}}{\Delta}\;\;\;\;\;0\;\;\;\;\;\;0\\
\\
\;\;0\;\;\;\;\;\;\;\;\;0\;\;\;\;\;\frac{1}{q_{22}}\;\;\;\;\;0\\
\\
\;\;0\;\;\;\;\;\;\;\;\;0\;\;\;\;\;\;0\;\;\;\;\;\frac{1}{q_{33}}\\
\end{array}
\right]
\end{equation}
where $q_{\mu\nu}=g_{\mu\nu}+bR_{\mu\nu}$, $\Delta=q_{00}q_{11}+
(\Phi^{'})^2$ and $\Phi=\sqrt{\kappa b}\phi$.
Taking the determinant of both sides of
(\ref{eq1}) gives
\begin{equation}
g=-q_{00}q_{11}q_{22}q_{33}\; .
\end{equation}
The field equations can then be written as
\begin{equation}
\left(P^{-1}\right)^{(\mu\nu)}=\sqrt{\frac{q_{00}q_{11}}{\Delta}}\;g^{\mu\nu}\;.
\end{equation}
The three independent equations are
\begin{equation}
q_{00}=-\sqrt{q_{00}q_{11}\Delta}\;\;g^{11}\; ,
\label{eq11}
\end{equation}
\begin{equation}
q_{11}=-\sqrt{q_{00}q_{11}\Delta}\;\;g^{00}\;,
\label{eq22}
\end{equation}
and
\begin{equation}
q_{22}=\sqrt{\frac{\Delta}{q_{00}q_{11}}}\;\;g_{22}\;.
\label{eq33}
\end{equation}
From (\ref{eq11}) and (\ref{eq22}) we find that
\begin{equation}
\Delta=g_{00}g_{11}\;\;\;\;\;\;\; and\;\;\;\;\;\;
g_{00}R_{11}=g_{11}R_{00}\;.
\end{equation}
Using these two equations gives 
\begin{equation}
g_{11}b^2R_{00}^2+2g_{00}g_{11}bR_{00}+g_{00}(\Phi^{'})^2=0\;.
\end{equation}
The solution to this equation is
\begin{equation}
R_{00}=-\frac{g_{00}}{b}\left[1-\sqrt{1-g^{00}g^{11}(\Phi^{'})^2}
\;\right]\;.
\end{equation}
Similarly we have
\begin{equation}
R_{11}=-\frac{g_{11}}{b}\left[1-\sqrt{1-g^{00}g^{11}(\Phi^{'})^2}
\;\right]
\end{equation}
and
\begin{equation}
R_{22}=-\frac{g_{22}}{b}\left[1-\sqrt{\frac{1}
{1-g^{00}g^{11}(\Phi^{'})^2}}\;\right]\; .
\end{equation}
The Einstein field equations can be written in covariant form as
\begin{equation}
G_{\mu\nu}=-\kappa \left\{\frac{F_{\mu}^{\;\;\;\alpha}F_{\nu\alpha}}{\sqrt{
1-\frac{1}{2}a^2F^2}}-\frac{g_{\mu\nu}}{a^2}\left[1-\sqrt{1-\frac{1}{2}a^2F^2}\;\right]\right\}\;,
\label{EFE}
\end{equation}
where I have taken $b=a^2/\kappa$ with $a$ constant and $F^2=F^{\alpha\beta}F_{\alpha\beta}$.
Note that this equation is not general. I have only shown that it is valid
if $P_{\mu\nu}$ is of the form given in (\ref{P}).
The energy-momentum tensor is identical
to the Born-Infeld energy-momentum tensor, except for the sign in the square
roots and the sign in front of the metric.

The electromagnetic field equations given in (\ref{eq22a}) can be written
as
\begin{equation}
\nabla_{\mu}\left[\frac{F^{\mu\nu}}{\sqrt{1-\frac{1}{2}a^2F^2}}\right]=0\;.
\label{EMFE}
\end{equation}
These field equations are the same as the Born-Infeld field equations with
the sign in the square root changed.

In Born-Infeld theory the square root $\sqrt{1-\frac{1}{2}a^2F^2}$, that
appears in the above equations, is replaced by $\sqrt{1+\frac{1}{2}a^2F^2}$. This difference is important in terms of
the allowed magnitude of the electric field. In flat space-time the Born-Infeld
square root is $\sqrt{1-a^2E^2}$, so that the maximum magnitude of the electric
field is $1/a$. In the theory presented here the square root is $\sqrt{1+a^2E^2}$,
so that the square root does not, by itself, constrain the magnitude of 
the electric field. For example consider the electromagnetic field equations
in a vacuum outside a spherically symmetric charge distribution in flat space-time.
The electromagnetic field equation is
\begin{equation}
\vec{\nabla}\cdot\left[\frac{\vec{E}}{\sqrt{1+a^2E^2}}\right]=0
\end{equation}
and the solution is
\begin{equation}
\frac{\vec{E}}{\sqrt{1+a^2E^2}}=\frac{Q}{r^2}\hat{r}\; ,
\end{equation}
where $Q$ is a constant. Note that although $E$ can be arbitrarily large
the left hand side in the above equation has a maximum magnitude of $1/a$
as $E\rightarrow\infty$.
Thus, we cannot let $r\rightarrow 0$ on the right hand side and the particle
has to have a finite size. In fact, any charged particle will have to have
a radius that is greater than or equal to $\sqrt{a|Q|}$.
  
Now consider the solution in curved space-time. The general spherically symmetric
solution for any theory with a Lagrangian of the form
\begin{equation}
-\frac{1}{2\kappa}R+L(F^2)\; ,
\end{equation}
is given by \cite{Ho1,Ho2,Pe1,De1,Wi1,Ol1} 
\begin{equation}
ds^2=-\left[1-\frac{2m(r)}{r}\right]dt^2+\left[1-\frac{2m(r)}{r}\right]^{-1}dr^2
+r^2d\Omega^2
\end{equation} 
and
\begin{equation}
D=\frac{Q}{r^2}dt\wedge dr\; ,
\end{equation}
where
\begin{equation}
D^{\mu\nu}=-2\frac{\partial L}{\partial F_{\mu\nu}}\;,
\end{equation}
\begin{equation}
\frac{dm(r)}{dr}=4\pi r^2H(s)\; ,
\label{BI}
\end{equation}
$s=-\frac{1}{4}D^{\mu\nu}D_{\mu\nu}=Q^2/(2r^4)$ and H is the ``Hamiltonian" given by
\begin{equation}
H=-\frac{1}{2}D^{\mu\nu}F_{\mu\nu}-L\;.
\end{equation}
The electromagnetic field equations are
\begin{equation}
\nabla_{\mu}D^{\mu\nu}=0
\end{equation}
and the energy-momentum tensor is 
\begin{equation}
T_{\mu\nu}=D_{\mu}^{\;\;\alpha}F_{\nu\alpha}+g_{\mu\nu}L
\end{equation}
The field equations (\ref{EFE}) and (\ref{EMFE})
can be obtained if we take
\begin{equation}
L=\frac{1}{a^2}\left[\sqrt{1-\frac{1}{2}a^2F^2}-1\right]\; .
\end{equation}
For this Lagrangian
\begin{equation}
D^{\mu\nu}=\frac{F^{\mu\nu}}{\sqrt{1-\frac{1}{2}a^2F^2}}\; ,
\end{equation}
\begin{equation}
H(s)=\frac{1}{a^2}\left[1-\sqrt{1-2a^2s}\right]
\end{equation}
and equation (\ref{BI}) can be written as
\begin{equation}
\frac{dm(r)}{dr}=\frac{4\pi}{a^2}\left[r^2-\sqrt{r^4-a^2Q^2}\right]\;.
\label{mprime}
\end{equation}
Thus, just as in the flat space-time case, the particle has to have a coordinate
radius that is greater than or equal to $\sqrt{a|Q|}$. 
  
To see if the space-time becomes singular as we approach $\sqrt{a|Q|}$
consider the Ricci tensor which is given by
\begin{equation}
R=-2\left[\frac{rm^{''}+2m^{'}}{r^2}\right]\;.
\end{equation}
From (\ref{mprime}) we have
\begin{equation}
R=-\frac{32\pi}{a^2}\left[1-\frac{r^2-\frac{a^2Q^2}{2r^2}}{\sqrt{r^4-a^2Q^2}}\right]\;.
\end{equation}
Thus, there is a curvature singularity at $r=\sqrt{a|Q|}$ and the particle
has to have a coordinate radius greater than $\sqrt{a|Q|}$ to avoid the singularity.

\section*{The Dirac Field}
In this section I will examine the Dirac field in Born-Infeld-Einstein theory.
To deal with spinors it is convenient to introduce a tetrad $e^a_{\;\;\mu}$
satisfying
\begin{equation}
e^a_{\;\;\mu}e^b_{\;\;\nu}\eta_{ab}=g_{\mu\nu}\;\;\;\;\;\;\;\; and
\;\;\;\;\;\;\;\; e^a_{\;\;\mu}e^b_{\;\;\nu}g^{\mu\nu}=\eta^{ab}\; .
\end{equation}
The Riemann tensor is defined to be 
\begin{equation}
R^{\;\;\;\;ab}_{\mu\nu}=\partial_{\mu}\omega_{\nu}^{\;\;ab}-\partial_{\nu}\omega_{\mu}
^{\;\;ab}+\omega_{\mu}^{\;\;ac}\omega_{\nu c}^{\;\;\;\;b}-\omega_{\nu}^{\;\;ac}
\omega_{\mu c}^{\;\;\;\;b}\; ,
\end{equation}
where $\omega_{\mu}^{\;\; ab}$ is the spin connection (see \cite{van1} for
a discussion of spinors)
  
The action will be taken to be
\begin{equation}
L=-\frac{1}{\kappa b}\left\{\sqrt{-det(g_{\mu\nu}+bR_{\mu\nu}+\kappa bM_{\mu\nu})}
-\sqrt{-det(g_{\mu\nu})}\right\}\; ,
\label{lagdirac}
\end{equation}
where $g_{\mu\nu}$ is to be expressed in terms of the tetrad,
\begin{equation}
R_{\mu\nu}=-R^{\;\;\;\;ab}_{\mu\beta}e_{a\nu}e^{\;\;\beta}_b\; ,
\end{equation}
$M_{\mu\nu}=-2i\bar{\Psi}\gamma_{\mu}D_{\nu}\Psi$, $\Psi$ is the Dirac 
field, $\gamma^{\mu}=e^{\;\;\mu}_a\gamma^a$,
$\;\{\gamma^a,\gamma^b\}=2\eta^{ab}$, $g_{\mu\nu}=e^a_{\;\;\mu}e^b_{\;\;\nu}\eta_{ab}$,
\begin{equation}
D_{\mu}\Psi_e=\partial_{\mu}\Psi_e+\frac{1}{2}\omega_{\mu}^{\;\;ab}\Sigma_{ab}\Psi_e\;,
\end{equation}
and $\Sigma_{ab}=\frac{1}{4}[\gamma^a,\gamma^b]$ are the generators of 
the Dirac representation. Note that the Dirac Lagrangian depends on the 
spin connection.
Of course, other choices for $M_{\mu\nu}$ are possible. For example
one could symmetrize on the indices $\mu$ and $\nu$. However, since $R_{\mu\nu}$
is not symmetric this does not simplify the field equations.
The Lagrangian, to lowest order, corresponding to this choice for $M_{\mu\nu}$
is given by (\ref{approxLag}) and is the usual Dirac Lagrangian.

Varying the action with respect to the field $\bar{\Psi}$ gives 
\begin{equation}
\left(P^{-1}\right)^{\mu\nu}\gamma_{\nu}D_{\mu}\Psi=0; .
\end{equation}
To lowest order we obtain the standard Dirac equation $\gamma^{\alpha}D_{\alpha}
\Psi=0$.  
Varying the action with respect to the tetrad gives, to lowest order, the
field equations
\begin{equation}
G_{\mu\nu}=i\kappa\bar{\Psi}\gamma_{\nu}D_{\nu}\Psi\; ,
\end{equation}
where the field equation, to lowest order, for $\Psi$ has been used.
  
The field equation for the connection is found by varying the action with
respect to the connection. Note that the connection appears in both the Ricci
tensor and in $M_{\mu\nu}$. The details are rather messy, but to lowest order
we get the usual result (see \cite{Vo2} for details). This is not 
surprising since to lowest order the
matter Lagrangian is the standard Dirac Lagrangian.

\section*{Mass Terms}
In this section I will consider the inclusion of mass terms into the theory.
Let $\tilde{M}_{\mu\nu}$ be the matter contribution without a mass 
terms. A simple approach to include a mass term would be to let
\begin{equation}
M_{\mu\nu}=\tilde{M}_{\mu\nu}+\Omega g_{\mu\nu}\; ,
\end{equation}
where $\Omega$ depends on the field variable. For example, in the case of
a scalar field $\Omega=\frac{1}{4}m^2\phi^2$ and for the Dirac field 
$\Omega=\frac{1}{4}m\bar{\Psi}\Psi$,
where $m$ is the mass of the field. For simplicity consider the case of the
scalar field. Equation (\ref{eq1}) gives the field equations
\begin{equation}
\sqrt{P}\left(P^{-1}\right)^{(\mu\nu)}=\frac{\sqrt{g}g^{\mu\nu}}{1+\frac{1}{4}\kappa
bm^2\phi^2}\;.
\end{equation}
The field equations given by varying the action with respect to the connection
are still given by (\ref{eq2}). Thus the connection is no longer the Christoffel
symbol with respect to the metric $g_{\mu\nu}$. In fact, the connection is
the Christoffel symbol with respect to the metric $h_{\mu\nu}=(1+\frac{1}{4}
\kappa bm^2\phi)^{-1}g_{\mu\nu}$. Thus, $R_{\mu\nu}$ and $P_{\mu\nu}$ will be symmetric.

One can also show that the Einstein equations are given by
\begin{equation}
R_{\mu\nu}=-\kappa\left[\nabla_{\mu}\phi\nabla_{\nu}\phi+\frac{1}{2}
m^2\phi^2\left(\frac{1+\frac{1}{8}
\kappa bm^2\phi^2}{1+\frac{1}{4}\kappa bm^2\phi^2}\right)g_{\mu\nu}\right]\;.
\end{equation}
These are not the standard Einstein equations with a mass term. 
However, if $\kappa bm^2\phi^2<<1$ we get the standard equations to 
lowest order. There
is nothing wrong with these field equations, but a simpler theory can
be found. This theory follows from the action
\begin{equation}
L=-\frac{1}{\kappa b}\left\{\sqrt{-det\left[g_{\mu\nu}+bR_{\mu\nu}+
\kappa b\left(\nabla_{\mu}\phi\nabla_{\nu}\phi+\frac{1}{2}m^2\phi^2g_{\mu\nu}\right)\right]}
-\sqrt{-det\left[\left(1+\frac{1}{2}\kappa bm^2\phi^2\right)^{\frac{1}{2}}g_{\mu\nu}
\right]}\right\}\; .
\label{mass}
\end{equation}
Varying this action with respect to the connection gives the same equations
as before and varying it with respect to the metric gives
\begin{equation}
\sqrt{P}\left(P^{-1}\right)^{(\mu\nu)}=\sqrt{g}g^{\mu\nu}\;.
\end{equation}
The connection is therefore given by the Christoffel symbol and 
the Einstein equations are
\begin{equation}
R_{\mu\nu}=-\kappa\left[\nabla_{\mu}\phi\nabla_{\nu}\phi+
\frac{1}{2}m^2\phi^2 g_{\mu\nu}\right]\;.
\end{equation}
Thus, the resulting field equations are exactly the Einstein equations
coupled to a massive scalar field.
  
A similar approach can be used for the Dirac and electromagnetic fields.
For the Dirac field the mass terms will be of the form 
$m^2\bar{\Psi}\Psi g_{\mu\nu}$ and for the electromagnetic field the 
mass terms will be of the form $m^2A_{\mu}A_{\nu}$. The electromagnetic 
mass term, $m^2A_{\mu}A_{\nu}$, is actually simpler to deal with than the
scalar field mass term since it is independent of the metric.

\section*{Conclusion}
The field equations for the Born-Infeld-Einstein action including matter
were derived using the first order or Palatini formalism. For the scalar
field it was shown that it is possible to chose a Lagrangian that
gives the standard Einstein equations minimally coupled to a scalar field
even in the massive case. For the electromagnetic and Dirac fields the standard
equations are only reproduced to lowest order.
  
The spherically symmetric electrovac equations were
studied in detail. It was shown that the resulting Einstein equations correspond to gravity coupled to a modified Born-Infeld theory.  It was also
shown that point charges are not allowed in the theory. All particles must have a finite size.

\end{document}